# Exploring the Potential of the Innate Immune System for Computers Network Security

Almotasem Bellah Alajlouni

**Information Technology Department, Al-Huson University College, Al-Balqa Applied University**
**Irbid, 21510, Jordan**

**Abstract**
The human body has a very effective Immune system used to protect the body from dangerous foreign pathogens. This paper aims at studying the immunology and understanding how it works, it also shaded light on the usage of the immunology principles in the computer network security. It also suggested a new network security model which detects attacks that invades the LANs. This study based on human immune system (IS). This model help protecting the datalink layer by suggesting solution to detect the foreign frames in computer network traffic. In this model, the frame format is changed in a way that prevents the sender from sending his MAC address, and he send a unique identifier (ID) instead. Moreover, a special network switch will replace the sender ID with the corresponding MAC address and forward the packets to their right destination.

## 1. Introduction

Our world is full with technology and smart devices, so it become gradually control our daily life and they all connect together to form our future life. Moreover, the number of computer attacks rise dramatically, in spite of governments' procedures to protect the internet and computer networks, hackers are not affected and they have unlimited access to any computer and data on the Internet [1]. However, human body has an amazing protecting system from foreign attacks, and this paradigm has to be applied to secure networks of the computer [2].

Human body has a very effective immune system which works carefully to protect it from dangerous the foreign pathogens (e.g. bacteria, viruses, parasites, and toxins). The main purpose of this study is to understand the immunology system of human body, in order to use its principles in the computer network security. The following section will be mainly discussing the following questions:
1. How does the immune system work?
2. How can it detect and defiance the foreign objects?
3. How can it recognize self from non self object?

"computer virus" word was firstly invented by Fred Cohen, (1984) to describe harmful codes, which propagates on computer network, and attacks other devices [3]. Thereafter, many antivirus systems were created, based on finding specific patterns of harmful codes. [4, 5].

Drawing on biological immune systems, several applications have been emerged. Kephart [6, 7] was the first one who introduce the concept of using human immunity defenses against computer viruses. Moreover, Forrest suggested a design of human immunity concept to be used in computer security systems [8]. He claimed that the main feature in any computer immune system must be self-distinctive from the others (non-self) [9]. So, to identify self-packets from non-self-packets he presented identification pattern from the source IP-address, the destination IP-address, and the destination port number.

This paper consists of five sections; the first is an introduction, the second talks about human Immune system, the third one discusses the proposed model of this research, the forth is the analysis of the that model, and the final section concludes the results of this research.

## 2. Human Immune System

Every protein or micro being has a unique genetic identity. Human body uses this genetic identity to know the class of this being, which is kind of protein or number of proteins found on top of the cellular wall of those micro beings. By using some a kind of receptors, the immune system determines if this being is an intruder or not [10].

All the cells of human body have a set of surface proteins that is used to mark them as self-tissue [10]. Generally, the immune cells do not attack the body tissues, that have the same set of self-markers. The major histocompatibility complex (MHC), which is group of genes that code for proteins found on the surfaces of cells used for self / Non-self-recognition [11]. Any cell that doesn't have this marker is identified as non-self and attacked [12]. Antigen is any foreign element may activate the response of the immune system in human body [13]. It may be a virus, a bacterium, foreign cell, or even just a piece of a foreign protein [14].

Generally, Immune system classified any unknown protein as an intruder, so it attacks that protein until it is recognized as a self-protein. In the same occurs in the surgeries of organs transplant; the immune system start attacking the new organ, so doctors give the patient an enzyme to slow down the response of the immune system toward the new organ from being attacked until it is recognized as harmless being [15].

The immunity system is classified into two classes; The innate immunity system and the adaptive immunity system [10]. The innate immunity system attacks antigens and pathogens equally through special type of cells called phagocytes. Those cells have surface receptors that recognize a wide spectrum of common bacteria [10].

On the other hand, the adaptive immunity system is more specific and specialized and it uses different types of cells called lymphocytes [10]. Those cells can only recognize specific antigen. This system may take several days or weeks to response to the effect of initial infection [13].

In this paper, the Innate Immunity system is our main concern, since it is responsible for determining intruders regardless of their class or type[10, 15]. In the regular conditions, the Innate Immunity system forms about 70% of the immune system cells in the body and it is mainly dependent on due its fast responding in discovering intruders[13]. In the case of discovering infection, this system Exceeds the adaptive Immunity System, which in its turn is responsible for determining the spread of the disease and developing the accurate cure for it (i.e. Adaptive IS needs between 6-8 days to response) [13].

The dedicated path for the immune system is known as lymph vessels These vessels extend to cover all the body. They also connected to blood vessels in special places called lymph node which is small bean-shaped node located along with the lymphatic vessels [10]. Each lymph node has a special room for immune cells to detect and attack antigens. The lymph nodes cluster is located in the neck, armpits, and abdomen [10].

## 3. The Proposed Model

Drawing on Immune system in human body of dealing with intruders or pathogens, we are going to propose a new immunity model for computer networks called Network Immune System (i.e. **NIS**). This model will take advantage of the Immune System of distinguishing foreign (non-self) messages from true (self) messages to protect the network from intruders. The main features of that mode as follows: Sending frames is not allowed by hosts unless they have unique identifiers (i.e. **ID**). Moreover, That ID has to be known to the connected switch in the network. Thus, the switch will discard any frame does not have that ID

In order to get specific ID, hosts have to register in the security server of the network by sending their information (e.g., IP address, MAC address, and username and password of the current user). Furthermore, the security server will approve registration process by sending ID to the intended host and consider it as a part of the network.

The links used for sending IDs should be limited to this purpose and not extend to the other normal data transfer through the network. When the system detects an intruder, the Adaptive immune system will be triggered and start classifying its defect type to dealing with it in the proper way.

3.1 NIS Model

Applying security model at transport-layer protocol (i.e. TLS protocol -Transport Layer Security- [16]), then, all applications use it will benefit the security services. Also, when security is provided at the network layer (i.e. IPsec protocol -IP Security- [17]), all above layers (transport and application layers) enjoy its security services[18, 19]. In the same manner, when security is provided at a datalink layer, then all the upper layers will exploit security services of the datalink. Therefore, this model is going to secure the datalink layer in Correspondence to the human Innate immune system.

In this model, the new switch (which is called Lymph switch) is similar to the Lymph Nodes in human body; it has to be connected to a main security server through a secure connection that is not used for data transfer. The security server (Which is called NIS server) uses those connections to send IDs to hosts.

Any frame reaches the lymph switch which does not have an ID or does have a fake one, will be sent to the NIS server to analyze its contents then to take a proper action. Procedures performed by the server may be simple like closing the port or tough by exciting the adaptive NIS to attack the intruder.

3.2 NIS Model Requirements

as shown in figure 1, NIS model needs to:
1. Modify the frame format by replacing the source and destination MAC addresses with the IDs.
2. Special secure switch (Lymph Switches); This switch has a special circuit to recognize the frame ID.
3. Main security server (NIS Server) works as a manager for the Artificial NIS.
4. Special communication links (Lymph link).
5. Protocol to register hosts to NIS Server.

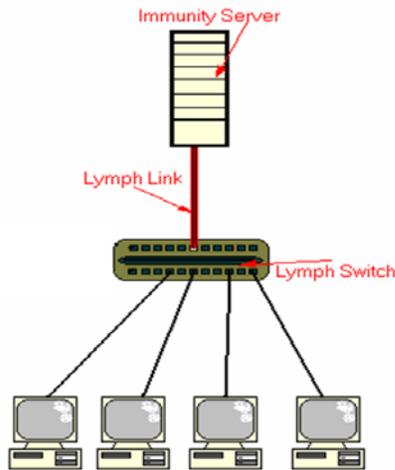

Figure 1: the proposed model requirements

Now we come to explain each of the above points.

### 3.2.1 Modify the frame format

Each datalink frame is sent through the network have source MAC address and destination MAC address. In this paper, the proposed model replaced the source MAC address with the computer identifier. So every frame is send from any computer must have the source identifier, the switch afterwards recognizes the identifier, if the identifier is valid the switch swaps it with the corresponding MAC address of the sender. Then replace the MAC address of destination with the identifier of destination. and pass it to its destination as shown in Figure 2.

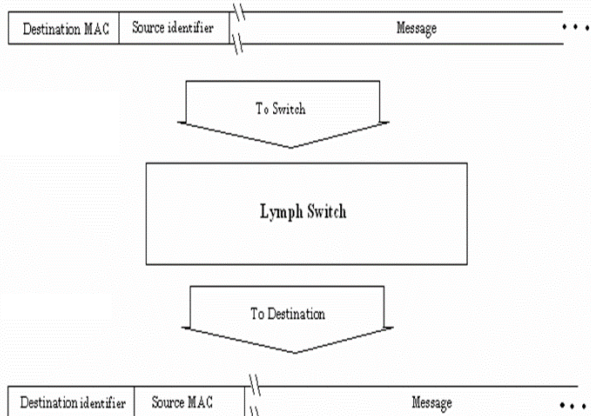

Figure 2: Replaced the source MAC address with the source ID as well as, destination MAC address with destination ID

The main idea from the previous swap is to secure the computer identifier from the sniffing, where every computer knows its ID only.

### 3.2.2 Lymph Switch

The switch ties each physical port with an ID of a host. This binding is done during process of registration. Any frame that cross through that port must have ID of that port or it will be discarded.

This switch has two modes; in the first mode it works as traditional switch with additional duties. It must check the frame sender's ID and swaps it with the corresponding MAC address. Also, it must check destination's MAC address and swaps it with the ID of destination.

To do so, each port in this switch must have Detector to recognize the ID. Also the switch Content Addressable Memory (CAM) table must have extra place to store the IDs. This table is used to map between IDs and MAC addresses. Moreover, in this mode the switch sends any illegal frame to immune server through special port. The job of the switch in this mode is illustrated in Figure 3.

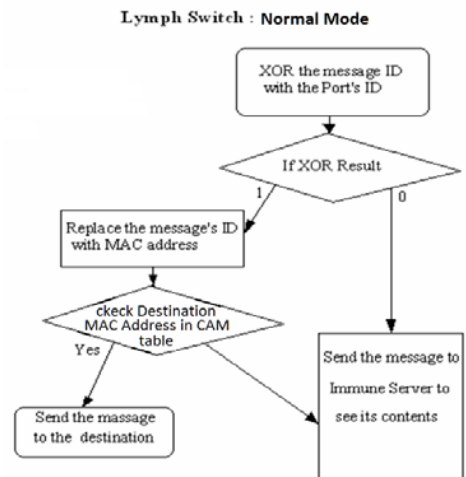

Figure 3: Lymph Switch in normal mode.

The second mode is authorization mode. In this mode the switch doesn't pass any message between hosts, except the authorization messages between NIS server and other computers. each host sends its encrypted information (IP address, MAC address, OS, user name and Password) to NIS server. the server processes this information and generates the ID and sends it back to the switch. The switch registers the MAC address and the ID of that host as well as the port in its CAM table then forwards the ID to the host.

### 3.2.3 NIS Server

This device is responsible for giving IDs to computers in the network and registering them on special database. It is connected to the Lymph switch through a special link that is not used for data transfer, for security issues. It contains a smart system to detect and attack the intruder. In another words, it is the holder of the Adaptive IS. This device must be hidden from hosts. So, it does not have IP address or MAC address.

3.2.4 NIS Protocol

In every period $T_p$ the NIS Server sends a registration request (REG_REQ) message to one of its Lymph switches and starts a timer for $T_r$. In this $T_r$, the NIS Server is waiting for response from hosts. after this time is finished no response message is accepted.

When the Lymph switch receives the REG_REQ message, it broadcasts it to all of the hosts connected to it. Also The Lymph switch changes its operation mode from normal mode to authorization mode. in this mode the switch does not allow any message to pass between the hosts until it receives a registration end (REG_END) message from the NIS Server.

When a host receives the REG_REQ message, it initiates a timer with a random number between 0 and 65535. In every period $T_d$ the host decrements the timer; $T_d$ must be less than $T_r/65535$. When the timer reaches to zero, the host sends a registration response (REG_RES) message to the NIS Server through its Lymph Switch. Then the NIS server generates the computer's ID and sends it to lymph switch. The lymph switch ties ID and MAC Address with port in its CAM table and passes the message back to the host.

After $T_r$ is finished the NIS Server sends message REG_END to lymph switch to go back to the normal mode. This process must be repeated every $T_p$, to remove the hosts that leave the network or to add new hosts and to prevents the ID hijack attack. The ratio between $T_r$ and $T_p$ must be large to keep the network performance in an acceptable value. NIS protocol is illustrated in Figure 4.

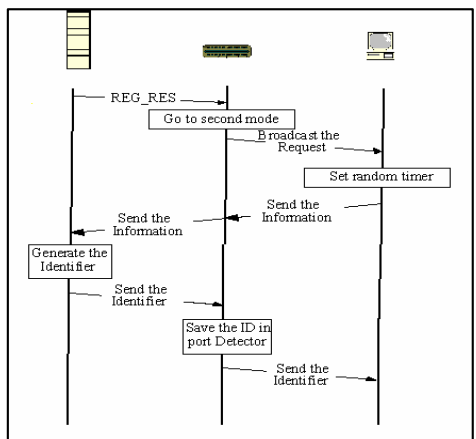

Figure 4: NIS Protocol

## 4. NIS Model Analysis

Many protocols have been established to secure application, transport, and network layers of a network. However, datalink layer is highly prone to several attacks, this is because of security lack in this layer. In this section, the most common attacks at datalink layer will be discussed. Also, the common scenarios of attacks and the response of NIS model on each attack will be address.

4.1 Common Attacks of Datalink Layer

Port Stealing attack exploits a switch capability to auto bind its ports to MAC addresses. The attacker uses the target host MAC address as source of ARP frames and floods the switch by these frames. The switch fancies that the target host on that port, and it forwards all traffic of the target host to that port, where the attacker ambushes his prize [20, 21].

In MAC Flooding attack, the attacker floods the switch with forged frames with faked source MAC addresses until its CAM (Content Addressable Memory) table is full. This leads to the switch broadcasts all traffic that do not have its address in CAM table. The attacker now able to sniff all frames in LAN [21].

Address Resolution Protocol (ARP) [22] is a stateless protocol used to get mapping between IP Address and MAC address. ARP spoofing or ARP cache poisoning is to send fake ARP reply that contains wrong IP-MAC addresses mapping. Each machine has a table to store the addresses mapping called ARP cache. This table is updated every time ARP reply is received, even though the ARP request has not been sent. This allow the attacker to pretend as legal host and then interrupt frames on the network. ARP spoofing is one of most risky attacks that suspicions LANs, that is because it is used as a base for other attacks such as man in the middle (MITM), denial of service (DoS), and session hijacking attacks [23].

In unsecure LAN the attacker can access the network simply by connects its host to any available port and starts its attacks. To secure this layer there is an intelligent switch that have secure port feature. In this feature every port is statically configured to associate with a specific MAC address. Although secure port feature succeeds to prevents Port Stealing and MAC Flooding attacks, this feature is not widely used because it is difficult to implement and maintained in large LAN [24].

The proposed model is designed to secure datalink layer against all of these attacks. it simulates the human innate immune system defense against different threats that encounter. In NIS model each port in lymph switch is tied with certain host that has unique ID to prevent Port Stealing attack. This ID is changed every period of time to prevent ID hijacking. This is done dynamically during process of authorization when lymph switch in mode one and that will secure the switch CAM table from MAC Flooding attack.

Any frame passes throw any port of the switch must have an ID that associated with that port. Hence, the lymph switch drops any frame that has incorrect ID and NIS server is notified. The NIS server analyzes all the dropped frames, and if the number of frames that were dropped in certain port exceeds a predefined limit, the NIS server notifies the lymph switch to shut down that port. This procedure is designated to detect and prevent ARP spoofing attack.

4.2 Common Scenarios of Attacks

To check NIS model, the attacks scenarios are addressed, although, prediction of all the scenarios are impossible. In fact, each time new security model is introduced, you will find someone trying to create scenario to break through that system. However, the common attacks scenarios are presented next, as well as, the response of NIS model on each scenario.

4.2.1 Scenario 1

In this scenario we will assume that the attacker has the ability to access the lymph switch and he tries to use empty port in that switch to do his attacks. NIS responses automatically and shutdown all the unused ports, so the attacker will not be able to send or receive any frame until the switch goes to mode one. In mode one, NIS server waits the authenticated information from the hosts to generate ID for each host. Thereafter, if the attacker fails to send a legal information (MAC address, IP address, username and password), his port will stay shut down when lymph switch goes back to mode two for another period of time.

4.2.2 Scenario 2

In this scenario, we will assume that the attacker steals the ID of another host and uses its port to access the network. If the attacker decides to apply the port stealing or MAC flooding attacks by using MAC address of other hosts, this model will help defeating his attack by replacing its MAC address with the ID. Therefore, he cannot use other MAC addresses as the source of his frames. Moreover, lymph switch CAM table will be built in mode one during process of authorization, by which the switch pends each port with the ID and MAC address of the authorized host connected on that port. Thereafter, any traffic in mode two will not be able to change the CAM table of the switch.

Moreover, If the attacker uses the stolen ID and tries to apply ARP cache poisoning attack by sending ARP replay with his MAC address, his APR replay will be delivered to victim host, but the attacker will not be able to get back any traffic because his MAC address is not found in lymph switch MAC table. Therefore, all the traffic that come from the victim host to the attacker will be discarded during the process of swapping the destination MAC address with the corresponding ID since it is not registered in the network.

4.2.3 Scenario 3

This is the most complicated scenario by which the attacker tries to steal the ID and MAC address of the authorized host to use it in ARP spoofing attack. As in scenario 2, the attacker has to access the network throw the port of that authorized host, then he will send it with MAC address of hijacked host to the target host. thereafter, the attacker will deceive the system for a period of time until the next registration process occurred, which is usually estimated with few minutes that is not enough to complete the intended attacks (e.g. a man in the middle (MITM) or denial of service (DoS) attacks).

the potential of this scenario to happen is rare; since it is difficult for the attacker to get the MAC address of the hijacked host, which is not found in frames that incoming to or outgoing from that host (the ID is used instead of MAC address).

## 5. Conclusions

This paper explored the potential of the innate immunity system for computer network security. It also presented new model for network security, which integrates defenses of network. Every micro being has a unique genetic identity. Human body uses this genetic identity to identify the class of this being. Micro being is defined as a kind of protein, or number of proteins found on the top of the cellular wall of those micro beings. Immune system can determine intruders by some sort of receptors.

The innate immunity system is responsible for identifying intruders with no regard to their class or type. In regular conditions, human body mainly depends on the innate immunity system, and this fact shows its quick response to discover intruders. Immunity system treats any foreign protein as intruder and starts to attack it.

This paper suggested a new immunity model in computer networks, which take advantage of the immune system for recognizing foreign (non-self) frames from true (self) frames to protect the network from intruders. This model helps securing the data link layer in correspondence to human innate immune system. Instead of sending its MAC address, the sender sends its unique identifier (ID). After the verification process has been occurred, lymph switch swaps sender's ID with the MAC address. It also swaps MAC address of receiver with receiver's ID then forwards frame to the destination.).

**Almotasem Bellah Alajlouni** is Full time lecturer at Information Technology Department in Al-Huson University College/Al-Balqa Applied University. Jordan. Eng. Alajlouni received the B.Sc. degree in Computers Engineering from Yarmouk University (YU), Jordan in 2003, and the M.Sc. in Computer Engineering from Jordan University of Science and Technology (JUST), Jordan in 2009. In the area of academic, Eng. Alajlouni is author of many publications in different research areas mainly nanoscale technology. He also has the 4th place in the contest of PATMOS "static Timing Analysis".